# Transitions in Microtubule C-termini Conformations as a Possible Dendritic Signaling Phenomenon

Avner Priel,* Jack A. Tuszynski* and Nancy J. Woolf[#]

*Department of Physics, University of Alberta Edmonton, AB, T6G 2J1, Canada
[#] Behavioral Neuroscience, Department of Psychology, University of California, Los Angeles, CA 90095-1563, USA

(Dated: May 11, 2005)

Keywords: neuron, synapse, MAP, ionic wave, protein filament, electrostatic interaction

We model the dynamical states of the C-termini of tubulin dimers that comprise neuronal microtubules. We use molecular dynamics and other computational tools to explore the time-dependent behavior of conformational states of a C-terminus of tubulin within a microtubule and assume that each C-terminus interacts via screened Coulomb forces with the surface of a tubulin dimer, with neighboring C-termini and also with any adjacent microtubule-associated protein 2 (MAP2). Each C-terminus can either bind to the tubulin surface via one of several positively charged regions or can be allowed to explore the space available in the solution surrounding the dimer. We find that the preferential orientation of each C-terminus is away from the tubulin surface but binding to the surface may also take place, albeit at a lower probability. The results of our model suggest that perturbations generated by the C-termini interactions with counter-ions surrounding a MAP2 may propagate over distances greater than those between adjacent microtubules. Thus, the MAP2 structure is able to act as a kind of biological wire (or a cable) transmitting local electrostatic perturbations resulting in ionic concentration gradients from one microtubule to another. We briefly discuss implications the current dynamic modeling may have on synaptic activation and potentiation.

# I. INTRODUCTION

Microtubules (MTs) have long been known to play key roles in the functioning of eukaryotic cells. Cell division, for example, is accomplished by the dynamic reorganization of the cytoskeleton and the segregation of the chromosomes by mitotic spindles that are composed of MTs (Alberts et al., 2002). Whereas all living cells contain MTs, in neurons they facilitate processes and interactions that extend over large (even macroscopic) distances. Elongated structural elements, for example the dendrites of large pyramidal cells of cerebral cortex, provide an environment where interactions that are electromagnetic in nature may propagate along the long axis of MTs aligned within them (Brown and Tuszynski, 1999).

Recently, it has become apparent that neurons utilize MTs in cognitive processing. Both kinesin and microtubule-associated protein 2 (MAP2) have been implicated in learning and memory (Wong et al., 2002; Khuchua et al., 2003; Woolf et al., 1999). Dendritic MTs in particular, are implicated in these processes, and it is highly probable that precisely coordinated transport of critical proteins and mRNAs to the post-synaptic density via kinesin along MT tracks in dendrites is necessary for learning, as well as for long-term potentiation (LTP) (Kiebler and DesGroseillers, 2000; Steward and Schuman, 2001). As shown schematically in Fig. 1, pre-synaptic axon terminals make synaptic contacts onto post-synaptic spines. Within the shaft of the dendrite, MTs act as tracks for kinesin's processive movement (Alberts et al., 2002); thus enabling them to transport proteins such as receptors. MTs also transport mRNA that translates to proteins critical to synaptic activity. In general actin participates in the dynamic changes of the dendrite and the formation of the spines. As for the interaction between the actin and MT as well as the possible role of actin in the signal processing, this is an interesting possibility but it is out of the scope of this paper and will be discussed elsewhere. Our computational analyses reported in the present paper focus on the C-termini of those MTs whose biophysical properties have a significant influence on the transport of material to activated synapses; as one example, cytoskeletal signal transduction and processing, and, as another, synapses undergoing LTP.

When Nogales et al. (1998) determined a 3D-structure of tubulin, the C-termini could not be resolved because the C-terminus region is highly flexible and lies outside the globular domain of tubulin. However, molecular dynamics enables one to model conformation dynamics of the C-termini using their known amino acid sequences in conjunction with the available crystallographic structures of tubulin. We show some relevant results in the next section.

The C-termini of tubulin are located on the outside of the MT and hence they interact strongly with other proteins, such as MAP2 and kinesin (Sackett, 1995). Indeed MAP2 binds to helices 11 and 12 near the C-termini of tubulin in MTs (Al-Bassam et al., 2002). While precise nature of the bond is not known at present, it is suspected to be ionic due to oppositely charged regions of tubulin and MAP2 in the binidng areas. To the best of our knowledge, a single MAP2 binds one tubulin dimer. The processivity of kinesin appears to critically involve an interaction with the C-termini of tubulin, although MAP2 may temporarily uncouple to avoid interfering with the movement of kinesin along the MT (Thorn et al., 2000). Removal of the C-terminus of tubulin profoundly affects kinesin processivity; in particular cleavage by subtilisin results in a four-fold decrease of the rate of kinesin transport (Wang and Sheetz, 2000).

In an effort to gain an insight into the role played by C-termini in the functioning of dendrites, we have developed a quantitative computational model based on the currently available biophysical and biochemical data regarding some key macromolecular structures involved (tubulin, its C-termini, and MAP2). Our model also accounts for the presence of ions in the surrounding aqueous environment. We start our modeling with the analysis of the possible conformations of the individual C-termini and calculate the resulting electric potential on the outside of the dimer. To elucidate the dynamic properties of C-termini and their interactions with MAP2 we study: (1) the relatively long-lived configurations of C-termini, (2) the interactions between C-termini and tubulin, (3) the interactions between nearest-neighbor C-termini, and (4) interactions between C-termini and MAP2 where we propose a model of wave propagation involving counter-ions that are weakly bound to MAP2.

## II. STRUCTURAL AND ELECTROSTATIC PROPERTIES OF TUBULIN INCLUDING C-TERMINI

In humans tubulin exists in dozens of homologous isoforms, some of which are preferentially expressed in the human brain (Lu et al., 1998). A 3D structure of tubulin was determined by Nogales et al. (1998) and later refined by Löwe et al. (2001) but several amino acids including some at the C-termini of the protein were not resolved, and hence these amino acid sequences were omitted from the (Protein Data Bank) PDB data files for tubulin (1TUB and 1JFF). However, we have created structure files that include the C-termini whose amino acid sequences correspond to those known for individual human tubulin isotypes. We have also performed computational experiments at physiological temperatures intended to find equilibrium conformations of tubulin's C-termini and these are shown later in the paper. The tubulin structure was a combination of 1JFF with a consistent set of structural elements from 1TUB for some regions absent in 1JFF. However, we must acknowledge the existence of serious limitations in the accuracy of computational studies of a protein as large and complex as tubulin.

An important physical property of proteins is their surface electric charge distributions, as these affect their interactions with other proteins and biomolecules. Fig. 2 illustrates the results of our visualizations based on the PDB file 1JFF. It shows a coarse-grained representation of the electric charge distribution on the tubulin dimer surface. The intention here is to show where charges of different signs are located, not necessarily how densely they are distributed at a given point. The fading of the font used represents the depth of view meant to express the 3D structure of the protein. This figure contains regions including the two C-termini that were not crystallographically resolved, but added here using relevant amino acid sequences. These C-terminal chains conformations are purely artifacts of the process by which they were added and should not be interpreted as corresponding to any real conformation. The figure and the potential calculation were in part produced using MolMol (Koradi et al., 1996).

The C-termini of tubulin are strongly negatively charged (having a net charge of up to 11 electrons each) and interact electrostatically with several other charged objects in their vicinity, namely: (a) the surface of the tubulin dimer (which is generally negatively charged, with as many as 16 negative charges per tubulin monomer, but which has a positively charged region(s) on its exposed surface that can bind a C-terminus), (b) with neighboring C-termini and (c) with adjacent proteins such as kinesin or MAP2.

Through the computer simulations that are later described in detail, we have found that the C-termini are likely to exist in two major conformational states: (a) a rather flexible conformation pointing away from the MT surface and subject to thermal fluctuations, or (b) a state in which the C-termini bind to the MT surface. Fig. 3 shows a tubulin dimer where two C-termini configurations are obtained from simulations (not described in this paper); in one of them both C-termini point upright away from the surface while the other dimer shows one upright C-termini and one bound. It is feasible that bound conformations of C-termini form a regular arrangement on the surface of a MT. Interestingly, a recent paper by Makrides et al. (2003) reports that MAP (tau) binding to neuronal MTs produces a collective change in the MT lattice structure. Their result obtained using AFM (atomic force microscopy) techniques is consistent with our prediction of the existence of a collective state of the C-termini that are bound to the tubulin surface. However, we deliberately avoided any relation to conformational changes of the dimers themselves. This is because the time scale for the collective motions of the C-termini is much shorter than that for the bulk of the protein's conformational motions. Besides, dimer conformational changes (if any) are much smaller with respect to the degrees of freedom available for the C-termini. While this section has dealt with the preliminary application of molecular dynamics simulations to individual tubulin monomers and dimers, in the next sections we intend to explore the possibility of interactions amongst the many C-termini of a microtubule as well as between the C-termini and MAPs. Our intention is to examine whether: (a) collective conformational states of these structural elements may be generated, and (b) whether these states can exhibit non-trivial dynamics, e.g. wave propagation. As will be elaborated in the discussion, such collective states may be reversibly switched on and off; this might be of significance to dendritic information processing and axonal trafficking. However, due to the complexity and computational demands of this problem, we are now faced with a need to simplify the modeling effort.

## III. MODELING THE C-TERMINI AS CHARGED RODS

Based on the assumption that C-termini dynamics plays an important role in the functioning of a dendrite, we develop a simplified model of a C-terminus and its interactions with neighboring structures. Analysis of the model indicates the existence of two metastable states that were observed in the full MD simulation performed on a single dimer, and allows further investigation of the interactions with MAP2.

The proposed model of the C-termini microtubule network is schematically illustrated in Fig. 4 where the tubulin dimer is considered to be the basic unit. Each dimer has two C-termini that may either extend outward from the surface of the protofilament or bind to it. The major goal here is to explore how MAP2 binding affects the states of the C-termini, which in turn may affect other sub-cellular processes, the processive motion of kinesin motor proteins transporting mRNA as they walk on the MT.

The most responsive structural elements of the system (i.e., elastic and electric degrees of freedom) can be best understood in terms of the conformational states of the C-termini. Each state of the unbound C-terminus is assumed to evolve such as to minimize the overall interaction energy of the system. Note that while the exposed surface of the dimer is, on average, highly negatively charged, it exhibits local regions of positive charge that attract the C-termini as thermal fluctuations (or directed fields) may cause them to bend into and bind in what we call a 'down' state. The energy difference between the two metastable states is relatively small, on the order of few $k_BT$ at room temperature, allowing for transitions between the two main conformational states to be thermally activated. We elaborate on this below.

In this preliminary analysis we map the interaction energy of a C-terminus with its nearest neighbors to investigate its static properties; dynamic properties will be visited later. Here, for simplicity, we model the C-terminus as a charged rod (of negligible width) that is attached at one end to the surface of the tubulin dimer, and can move otherwise above the surface as a rigid body (see Fig. 5). We describe its location in terms of spherical polar coordinates as: ($\theta$, $\phi$) (azimuth, elevation). The main contributions to the interaction energy of the test C-terminus are: 1) its nearest neighbors' C-termini, 2) the negative surface of tubulin dimer below and 3) the positive regions on the surface. Hence, we can describe the total energy of the $i$th test C-terminus as follows:

$$E_i^{tot} = \sum_{j \in nn} E_{i,j}^{cc} + E_i^{cd} + E_i^{cp} \qquad (1)$$

The different contributions to the interaction energy are now described explicitly. We use in our calculations the concept of Debye screening due to the presence of ions in a solution (Daune, 1999). The electrostatic interaction energy between two C-termini labeled ($i$, $j$) can be approximated as:

$$E_{i,j}^{cc}(\theta_{i,j}, \phi_{i,j}) = K\lambda^2 \int_0^L \int_0^L dl_i dl_j \frac{e^{-(D(\bullet)/\lambda_D)}}{D(\bullet)\varepsilon_r(D(\bullet))}$$

$$D(l_{i,j}, \theta_{i,j}, \phi_{i,j})^2 = (l_i \cos\phi_i \cos\theta_i + x_i - l_j \cos\phi_j \cos\theta_j - x_j)^2 +$$
$$(l_i \cos\phi_i \sin\theta_i + y_i - l_j \cos\phi_j \sin\theta_j - y_j)^2 + (l_i \sin\phi_i - l_j \sin\phi_j)^2 \qquad (2)$$

where ($x_{i(j)}, y_{i(j)}$) is the (fixed) binding location of the C-termini to the surface, $\theta_{i,j}$ is an abbreviation for ($\theta_i$, $\theta_j$), $D(\bullet)$ stands for the distance function, $L$ is the length of the C-terminus, $\lambda$ is the linear charge density, $K=(4\pi\varepsilon_0)^{-1}$, $\varepsilon_r$ is the relative dielectric constant (which for short ranges in aqueous solution is 1–80), and $\lambda_D$ is the Debye screening length that depends on the ions and their concentrations in the surrounding solution.

The tubulin dimer has a complex surface both geometrically and electrostatically. For simplicity, we model it as a 2D rectangular area with a certain average charge density $\sigma$ with several localized positively charged regions whose locations are based on calculations described above, see Fig. 2. Considering the dimer and one of its C-termini, with the center of the dimer at the origin, the negative surface contribution to the interaction energy is given by:

$$E_i^{cd}(\theta,\phi) = K\lambda\sigma \int_0^L \int_{-\frac{W}{2}}^{\frac{W}{2}} \int_{-\frac{H}{2}}^{\frac{H}{2}} dl\,dw\,dh \frac{e^{-(D(\bullet)/\lambda_D)}}{D(\bullet)\varepsilon_r(D(\bullet))} \quad (3)$$

$$D(l,w,h,\theta,\phi)^2 = (l\cos\phi\cos\theta - h - x_0)^2 + (l\cos\phi\sin\theta - w - y_0)^2 + (l\sin\phi + \Delta)^2$$

where $(x_0, y_0)$ is the binding location to the surface of the C-terminus, $H$, and $W$ are the height and width of the dimer, respectively, and $\Delta$ is the distance in the $z$-direction separating the C-terminus and the dimer surface. Similarly, the contribution due to positively charged regions (points) is given by:

$$E_i^{cp}(\theta,\phi) = K\lambda q_e \sum_{j=1}^{N_p} \int_0^L dl \frac{e^{-(D_j(\bullet)/\lambda_D)}}{D_j(\bullet)\varepsilon_r(D_j(\bullet))} \quad (4)$$

$$D_j(l,\theta,\phi)^2 = (l\cos\phi\cos\theta - x_j - x_0)^2 + (l\cos\phi\sin\theta - y_j - y_0)^2 + (l\sin\phi + \delta)^2$$

where $N_p$ is the number of positive charges on the dimer surface, $\delta$ is a minimum $z$-displacement of a positive region with respect to the surface (the residue is, to some extent, exposed) and $q_e$ is the electron charge.

Based on the above contributions, we calculated numerically the interaction energy surface as experienced by a C-terminus. The main assumptions behind this analysis are:

- The empirical estimate of the relative dielectric constant $\varepsilon_r$ (Mallik et al., 2002) is linearly approximated by:

$$\varepsilon_r(d) = \begin{cases} 1 + \frac{79 \cdot d}{1.7} & d \leq 1.7 \text{ nm} \\ 80 & d > 1.7 \text{ nm} \end{cases} \quad (5)$$

- Further effects of the ionic solution are taken into account only via the inclusion of the Debye-Huckel factors (Daune, 1999).

The governing equations were evaluated on a $\theta$–$\phi$ grid defining the inclination of the C-terminus. Fig. 5 depicts the simulation setup where a test C-terminus is rotating with two degrees of freedom (the base is fixed), ($\theta$, $\phi$) denoting the azimuth and elevation, respectively, and the rest of the environment is frozen, i.e. neighboring C-termini, and the dimer's surface, a negatively charged area with specific positive regions. The model parameters used for the numerical evaluation are given as follows. First, we have used the geometrical dimensions of the tubulin dimer as: $H$=8 nm, $W$=6 nm. Then, we assumed that approximately 2/3 of the net average charge of tubulin is exposed on the outer surface, which results in the surface charge density of: $\sigma = 12e^- \cdot (4 \text{ nm})^{-2}$. Next, a reasonable value of the linear charge density for the C-termini: $\lambda = 10e^-/L$ (where $L$ is the C-terminus length, $L=3.5$ nm). The values of the displacement parameters have been estimated at $\Delta = 4$ Å, $\delta=1$ Å.

Figure 6 depicts the interaction energy for several elevation cuts (angles are given in the figure) at azimuth angles, whereas Fig. 7 shows a full surface plot. It transpires that the 'up-state' has the lowest energy (it corresponds to the C-terminus being perpendicular to the tubulin's surface). However, the cone-angle created by the constraint $E-E_0 < 50$ meV (where $E_0=E\ (\phi=90^0)$) is about $40^0$. This means that the C-termini can move readily within this cone due to thermal fluctuations ($k_BT$ is approximately 25 meV at physiological temperatures). But, an important result is the existence of a local minima associated with the 'down-state' which is only 100 meV higher than the straight-up state, see Fig. 7. A saddle point between the two states is ~160 meV above the 'up-state'. What we deduce here is that, at least under the assumptions of the model, there are two metastable states with a moderate energy barrier separating them. The insert in Fig. 7 shows a close-up near the saddle point.

Finally, we wish to discuss ion condensation that may be expected to take place on MAPs and the C-termini of MTs due to their strong negative charge exposure to the aqueous environment. Many years ago, Manning (1978) postulated an elegant theory that poly-electrolytes may "condense" neutralizing ions from their surroundings. According to Manning's hypothesis counter-ions condense along the polymer, if a sufficiently high linear charge density is present on the polymer's surface. Thus, a linear polymer may be surrounded by counter-ions from the saline solution such that co-ions of the salt solution are repelled so that an ion depleted volume is created. The sum of surface charges and associated counter-ions depends on the counter-ion's valences and Bjerrum lengths. In general, depending on the charge density and the geometry of the charged surface exposed to ionic solutions (quasi-linear, cylindrical, spherical, etc) some of the ions condense on the surface binding very strongly to the opposite charges on the surface, while a number of ions remain fairly free to move about in the solution still being attracted to the surface. The proportion of the condensed ions to the surface charge can be evaluated for a given situation. The Bjerrum parameter is a phenomenological property of the ion's ability to compensate for the surface charges, and depends on temperature and the solvent. In 1926, Bjerrum developed a modification of Debye-Hückel ion-pair formation theory provided that the ions are small, of high valence, and that the dielectric constant of the solvent is small. The Bjerrum length is the distance at which the Coulomb energy of the screened charges equals $k_BT$, the thermal energy, creating an equilibrium from which the charges do not preferentially move. The cylindrical volume of ionic depletion outside the ion cloud surrounding the polymer serves as an electrical shield. Thus, the 'wire-like' behavior of such a structure is supported by the polymer itself, and the adsorbed counter-ions. For all practical purposes these are bound to the polymer. The strength of this interaction is such that even under infinite dilution counter-ions are still attracted to the polymer and do not diffuse away from the polymer. Although originally postulated for such poly-electrolytes as DNA, this theory applies to highly charged one-dimensional polymers such as the C-termini of microtubules or MAPs. In recent experiments (Stracke et al., 2002) it was demonstrated under the influence of externally applied electric fields that microtubules drift in the field direction, indicating that not all the protein charge is compensated by bound counter-ions.

# IV. DETAILED DESCRIPTION OF THE C-TERMINAL MODEL

In this section we extend our model in order to simulate the dynamic behavior of a C-terminus. To enable exhaustive simulations, we model the C-terminus as a sequence of beads with flexible connections, instead of the rigid rod, in what is known as the bead-spring model, a common method for simulating polymers. The simulation takes into account the following types of interactions:

A.  <u>Electric field of the dimer</u>.

Since the tubulin surface is complicated we use a dense grid that models its charge distribution. The outer surface can be characterized as mainly negative with a few small positive pockets (by visual inspection of our earlier calculations we estimate the existence of 7 such pockets per dimer). The computation of the interaction between the C-terminus and the dimer's surface is highly intensive if done for each update of the position. For example, if we use a 2Å resolution, the number of grid points is $(8/0.2)*(6/0.2)=1200$, assuming a 6 x 8 $nm^2$ surface area. If we use a 10-bead model then the number of point-to-point interactions becomes 12,000. A reasonable number of updates is $10^5$ to $10^6$, so this type of interaction becomes a serious bottleneck. To overcome this problem, we assume that the surface is static and evaluate the field it exerts in a box of (6 x 8 x 6) $nm^3$ volume whose base is the surface. This can be done even in a finer grid than described earlier. Once we have the (static) field vector at the box grid points, we apply real-time linear interpolation for each update to obtain the field at the position of each bead. We assume a negative charge of $-12e^-$ per tubulin monomer surface, and $+7e^-$ in positive pockets per dimer, hence the net negative contribution to the surface charge distribution is taken to be (-12-7/2) per monomer.

B.  The <u>Bead-spring model.</u>

A C-terminus has a molecular weight of ~2kD and is arranged as a long flexible tail. Instead of trying to model its entire secondary structure, we use an approximation where 1–2 amino-acids are modeled by a bead with an equivalent mass while the charge is distributed uniformly among the beads. To maintain the chain we use the following interactions:

**Coulomb interactions** between the electric charges;

- **The Lennard-Jones potential**: short range forces between all beads whose distance is shorter than a cut-off. This force is responsible for the strong short-range repulsion. The attractive part of this potential is truncated. The equation used is given below:

$F_{LJ} = 24\varepsilon\, r[2(\sigma/r)^{12} - (\sigma/r)^6]$ if $r < R_c$ (6a)

$F_{LJ} = 0$ if $r > R_c$ (6b)

- **FENE**: Finite-Extensible-Nonlinear-Elastic springs forces that enforce bond connectivity along the chain. This force is calculated between adjacent beads only, referred to as nearest neighbors (nn).

$F_{fene}(r) = KR_0^2 r/(R_0^2 - r^2)$ if $r < R_0$ (7)

Note that a singularity exists at $r = R_0$.

- **Angular forces**: These are the two types of forces that restrict the angle between every three adjacent beads and the planar angle involving every four adjacent beads, i.e., harmonic angular potential and harmonic dihedral potential.

*The harmonic cosine potential*: (for simplicity we give here the potential and not the force):

$U_{angle} = K\theta a\, (\cos(\theta) - \cos(\theta_0))^2$ (8)

where $\theta$ is the angle between every three adjacent beads.

*The harmonic dihedral potential*:

$U_{dihedral} = K\theta d(1+\cos(n\varphi - \varphi_0))$ (9)

where for the quartet (*ijkl*) we define φ = (*ijk*) ∠ (*jkl*), the angle between the plane spanned by (*ijk*) and the plane spanned by (*jkl*).

The other two contributions to the equation of motion are:

- **Friction forces**

$$F_f = -m\Gamma v, \tag{10}$$

  where $v$ is the velocity vector and $\Gamma$ is the friction coefficient.

- **Random forces** representing thermal fluctuations denoted by $W(t)$ and satisfying

$$\langle W(t) \rangle = 0, \quad \langle W_i(t)\, W_j(t') \rangle = 6\, k_B T\, m\Gamma\, \delta_{ij}\, \delta(t-t') \tag{11}$$

This combination of friction and noise terms is described in the fluctuation-dissipation theorem and often referred to as Langevin dynamics. The initial conditions adopted in our simulations had C-termini in a straight conformation normal to the surface. We used 600 runs that lasted 500 ps each and they took up 400 ps for relaxation and 100 ps for stabilization of the various resultant conformations that were then statistically analyzed. A detailed description of the model parameters chosen for our simulations is given in Appendix A. Note that the external force is taken in the x-direction with a random projection [–0.2..0.2] in the y-direction. What our simulation indicates here is the ability of an ionic wave to trigger a corresponding wave of C-termini state changes from their upright to downward orientations (see Fig.8). In the next section we address the related question whether or not an ionic wave can propagate along a MAP2 that interconnects two neighboring MTs. We show views of four C-termini states produced used this model in Fig. 9. Each has a different conformation with examples of the 'up' and 'down' states evident, as are two more complex states each with partial binding to the tubulin surface.

In Figs. 10 and 11 we have presented the results of our calculations of the energy minimized positions of the individual beads representing the amino acids of the C-termini in two equivalent forms. The distributions shown in Fig. 10 describe the minimal energy positions of beads representing the constituent residues of a C-terminus ranging from 3 to the end for each run after relaxation. The energy minimized value can come from each of the beads at positions from 3 to 11. The first bead is permanently connected to the surface and hence has no freedom to move. The surface is taken to be at the vertical coordinate $z = 0.45$ nm for technical reasons. Fig. 11 shows the cumulative sum of the probabilities corresponding to the data shown in Fig. 10. It is shown that the probability of the down position which includes all cases of full or partial attachment is ~15%.

## V. MODEL OF AN IONIC WAVE PROPAGATION ALONG MAP2

The next step in our investigation focuses on the characteristics of MAP2 and its ability to function as a 'wave-guide' that transfers the conformational change in a C-terminus state to an adjacent MT. Following the description of the MAP2 and its electrostatic properties we describe a simplified model of the interaction between MAP2 and its ionic environment via counter-ions and their role in propagating the perturbation exerted by an adjacent C-terminus.

The 3D structure of MAP2 is unknown. However, MAP2 binding with MTs has been investigated. The binding region is located at the C-terminus and the bond between MAP2 and a MT appears to be electrostatic in nature. Pedrotti et al. (1994) showed that MAP2 binds to MTs in a concentration-dependent manner. Once bound, a high NaCl concentration is not effective in reducing the concentration of MAP2 bound to MTs. During binding $MgCl_2$ is 10 times as effective as NaCl at decreasing MAP2 binding to MTs. The above authors concluded that MAP2 may "screen" electrostatic repulsion between tubulin dimers.

Our estimates of charge on the surface of MAP2 are based on a published sequence (Tokuraku et al., 1999) in the Swiss-Prot database (entry P11137: MAP2_HUMAN). We found that the binding domain is composed of roughly the first 418 amino acids so we take the remaining 1410 amino acids of MAP2b to constitute the projecting domain. Assigning a single positive charge to each Lys and Arg residue, and a negative one to Asp and Glu with His being given an average value of +1/2 due to its isoelectric point and leaving the remaining residues neutral, we take the total negative charge of the projecting domain to be –150 e⁻ and its charge distribution to be very close to uniform.

We are now able extend our model to include MAP2 as a supporting structure (backbone) to which counter-ions are attached. We are aware of the still unresolved controversy regarding the characteristics of the binding between the projecting domain of MAP2 and an adjacent MT. While it is still to be determined if a single MAP2 makes direct physical contact with a MT, it is apparent to us that it at least reaches the immediate vicinity of a given MT. However, our model's results presented below are unaffected by the type of binding present as long as the distance between these two structures is on the order of 1 nm. To simplify the model we assume that the counter-ion attracting sites are equidistant and arranged in a straight chain along the MAP2. We envisage that the counter-ions move only in a plane perpendicular to a cylinder representing the MAP2 (see Fig. 12). We show that a perturbation applied to the counter-ions at one end of the MAP2 will drive them out of equilibrium and initiate a wave that travels along the MAP2. Subsequently, we derive the equations of motion and investigate the properties of the traveling wave that emerges.

To facilitate the implementation of the model we use the harmonic approximation and represent the chain as a 2D-spring system in which each counter-ion is strongly bound to a MAP2 site and weakly bound to its two nearest-neighbor counter-ions. The main justification for taking only the nearest-neighbor interaction into account is the strong screening effect imposed by the solution (see the previous section for discussion). In addition, we assume external forces of two kinds; the damping viscous force due to temperature fluctuations and a perturbation exerted by a C-terminus while changing its state (details given below). An additional constraint is provided by reflecting boundaries located approximately 0.2 nm from the MAP2 (representing a repelling force).

Our main focus in the following investigation is the characterization of the propagating perturbation along the MAP2. The results reported below were obtained from MD simulations of this model. The model Hamiltonian can be written as follows:

$$H = \frac{M}{2}\sum_i \dot{\mathbf{r}}_i^2 + \frac{1}{2}K^b \sum_i (|\mathbf{r}_i - \mathbf{b}_i| - b_0)^2 + \frac{1}{2}K^c \sum_i (|\mathbf{r}_i - \mathbf{r}_{i-1}| - l_0)^2 + W(\mathbf{r}) \qquad (12)$$

where we have assumed a harmonic potential for the electrostatic interaction with $K^b$ and $K^c$ denoting the force constants of the MAP2-counterion and counterion-counterion interactions, respectively, $\mathbf{r}_i$ is the position of the counter-ion, $\mathbf{b}_i$ is the $i$th MAP2 binding site, $b_0$, $l_0$ are the equilibrium chain distances, $M$ is the counter-ion's mass and $W(\mathbf{r})$ describes the site dependent potential (which for the purpose of this calculation is ignored). Using the following notation, we obtain

$$\mathbf{u}_i^{CM} = \mathbf{r}_i - \mathbf{b}_i, \quad u_i^{CM} = \|\mathbf{u}_i^{CM}\|, \quad \hat{\mathbf{u}}_i^{CM} = \frac{\mathbf{u}_i^{CM}}{u_i^{CM}},$$

$$\mathbf{u}_i^{CC} = \mathbf{r}_i - \mathbf{r}_{i-1}, \quad u_i^{CC} = \|\mathbf{u}_i^{CC}\|, \quad \hat{\mathbf{u}}_i^{CC} = \frac{\mathbf{u}_i^{CC}}{u_i^{CC}}, \qquad (13)$$

where the symbols *CM* and *CC* label 'counterion-MAP' and 'counterion-counterion' interactions, respectively. The corresponding equations of motion for the counter-ions are given as follows:

$$M\ddot{\mathbf{r}}_i = -K^b(u_i^{CM} - b_0)\hat{\mathbf{u}}_i^{CM} - K^C\left[(u_i^{CC} - l_0)\hat{\mathbf{u}}_i^{CC} - (u_{i+1}^{CC} - l_0)\hat{\mathbf{u}}_{i+1}^{CC}\right] - \gamma \dot{\mathbf{r}}_i + \mathbf{f}_i \quad (14)$$

where damping and the external force, *f*, have been explicitly introduced. In order to solve these equations, we applied the velocity-Verlet algorithm for time integration (Swope et al., 1982).

Let us now describe the rationale and assumptions underlying the choice of parameters in the model, $K^{b(c)}$, $b_0$, $l_0$, $M$ and $\gamma$. The equilibrium distance between counter-ions is dictated by our assumption about the average distance between binding sites. As discussed above, from the analysis of the residues of MAP2 we infer that there are approximately 150 electronic charges. Assuming that MAP2 is a 50 nm rod-like structure, we obtain a charge distribution of about 3e⁻ per nanometer along the MAP2 axis (length estimation is obtained from average measurements of the MAP2). Since our model actually represents a two-dimensional structure perpendicular to the MAP2 surface, we assume the average charge separation to be $l_0$= 1 nm. The bond length for typical counter-ions in ionic solution is 2.5-4 Å; in our simulations we, therefore, used $b_0$=3.5 Å. The representative mass of the counter-ion chosen here is an average between sodium and calcium ions, *M*=30 *amu*, or approximately 5×10⁻²⁶ kg. The damping factor γ has units of *amu · ps⁻¹*; in our simulations we have used γ=1 *amu · ps⁻¹* which is typical in molecular dynamics simulations. The MAP2-counterion bond force constant takes a typical value of $K_b$=10 *kcal·mol⁻¹ ·Å⁻²* and the counterion-counterion force constant is on the order of $K_c$=0.3 *kcal·mol⁻¹ ·Å⁻²*. The former estimate is comparable to the strength and distance involved in the formation of a monovalent ion's bond with a protein. The choice of the second force constant reflects a weak Coulomb interaction between adjacent counter-ions. At any rate, it is the ratio of these two constants that is more important in our calculations than their absolute values.

The interaction between the MAP2 and the C-terminus actually includes a layer of counter-ions surrounding each of these structures. In our model the C-terminus is located 4 nm from the MAP2-base and is initialized at a given value of ($\theta,\phi$). It then falls at a certain angular velocity until it reaches the tubulin surface. We apply MD updating steps to the counter-ions only, since the C-terminus in this case merely serves as a source of perturbation. In parallel, we evaluate the properties of the wave generated by this perturbation. A typical initial orientation is $\theta$=180⁰, $\phi$=50⁰ and the C-terminus collapses over a characteristic time of 10 ps.

The following figures show several properties of the wave propagation along the chain of $N=50$ counter-ions. Figure 13 depicts the counterion displacement parallel to the MAP2, $u_i$, where $i$ denotes the $i$th counter-ion. The counter-ions near the MAP2-base are attracted by the C-terminus and the displacement is effectively negative. The wave propagates along the MAP2 and then is reflected back from the other end; this process may re-occur a few times. The amplitude of the reflected wave is attenuated and depends on the value of the damping factor $\gamma$. The asymmetry between the two edges is mainly an artifact of the experimental setup in which the C-terminus is located only at one end; hence the other end has fewer constraints and exhibits more fluctuations. The next two figures, Fig.14 and Fig. 15, present the first derivatives of the displacement with respect to the time and space, respectively. More specifically, the temporal and the spatial derivatives are estimated from the simulation updating steps as follows:

$$\frac{du_i(t)}{dt} \approx \frac{u_i(t+\Delta t) - u_i(t)}{\Delta t} \quad (15a) \qquad \frac{du_i(t)}{dy} \approx \frac{u_{i+1}(t) - u_i(t)}{\Delta y_i} \quad (15b)$$

where $\Delta t$ is the simulation time step, $\Delta y_i$ is the distance between two adjacent counter-ions and the spatial derivative is obtained only in the y-direction (parallel to the MAP2 axis).

The profiles of the displacements and their derivatives reveal the following interesting result. The projection of the initial perturbation along the MAP2 propagates almost as a "kink", mainly due to the large $K_b/K_c$ ratio; only after the second reflection can one observe a significant spread of the packet. Determination of the phase velocity is done by evaluating the following quantity for all sites:

$$\{t_i : u_i(t) = u_0 \quad , \quad u_i(0) = 0\} \tag{16}$$

i.e., we find the set of time values $\{t_i\}$ for which the displacement of the $i$'th counter-ion is $u_0$ for the first time. We measure the first passage time for all sites from which the propagation velocity is evaluated. Figure 16 shows the results for the above parameters with a velocity estimation of $v_{\text{ph}} \approx 2 \, nm \cdot ps^{-1}$. Finally, it is worth mentioning that the velocity scales with the square root of $K_c$. More details of this analysis will be published separately.

## V. SUMMARY AND DISCUSSION

In this paper we have modeled static and dynamic properties of the C-termini of tubulin dimers that comprise neuronal microtubules. We have used various computational tools including molecular dynamics in order to explore the dynamic behavior of conformational states of a C-terminus of tubulin within a given microtubule and assumed that each C-terminus interacts via screened Coulomb forces with the surface of a tubulin dimer, with neighboring C-termini and also adjacent MAP2s. Each C-terminus was shown to either bind to the tubulin surface via one of the several regions we identified as positively charged or to explore the space available in the solution surrounding the dimer fluctuating above the tubulin surface with higher probabilities for conformations with a larger inclination angle. We found that the preferential orientation of each C-terminus is away from the tubulin surface but binding to the surface may also take place at a considerably lower probability. The static properties of the C-terminus where studied by analyzing the energy surface of a system composed of a test C-terminus modeled by a charged rod interacting with the tubulin dimers below and the neighboring C-termini. This analysis confirmed the hypothesis of two major states where the volume described by cone-angles above 50 degrees has energy differences of ~2KT at room temperature; we also found a clear barrier of ~6KT between the two major types of states. We further evaluated the statistical properties of the two major states by studying the dynamic properties of a polymer-like bead-spring model of the C-terminus. The results indicate that the system has indeed two major states with a strong bias towards the stretched-up state. In other words, the system tends to be mostly in the 'up-state' unless driven towards one of the surface binding sites where the probability of attraction increases. The statistical bias favoring the up-state versus the down state of C termini may be tipped depending on the actual physiological conditions, in particular the presence of local polarizing fields and/or screening effects. This may be affected by many processes, for example the effect of actin polymerization in the vicinity of a given microtubule or even by ordering of water molecules surrounding the MT surface.

Finally, we studied the properties of ionic wave propagation from one microtubule to another along MAP2. The trigger for this wave may be a near by C-terminus changing state or ionic flux, e.g., from an adjacent ion-channel. Under the non-condensed counter-ions assumption we derived the equation of motion for a one-dimensional chain of counter-ions along the MAP2 main axis and evaluated the properties of such a wave. It has been found that the wave may propagate at a speed of ~2nm/ps (=2000 m/s) with a small amplitude attenuation. The results of our model suggest that perturbations generated by the C-termini interactions with counter-ions surrounding MAP2 may propagate over distances greater than those between adjacent microtubules. Interestingly, a recent paper by Ripoll et al. (2004) has proposed an important integrative role for ion waves guided by protein polymers in providing a new means of signal transduction in eukaryotic cells. Our model offers a quantitative description of such a possibility in neurons.

Below, we briefly discuss implications the current dynamic modeling may have on synaptic activation and potentiation. The results of the molecular dynamics modeling performed here raise the possibility of transmitting electrostatic perturbations collectively among neighboring C-termini and from C-termini located on one MT to those C-termini on another MT via a MAP2 that physically interconnects MT's in dendrites. Collective conformational states of C-termini and a transmission of perturbations among the C-termini were shown to be a viable mode of behavior in dendritic microtubules. This may suggest the possibility of a hitherto unexplored information processing mechanism operating at a sub-neuronal level. For example, MAP2 and kinesin bind near to the C-termini on tubulin and the conformational states of C-termini affect this binding (Al-Bassam et al., 2002; Thorn et al., 2000). Hence these conformational states of the C-termini must at some point be taken into account in order to understand neural processing that depends on transport of synaptic proteins inside of neurons (see Fig. 1). In this connection, Kim and Lisman (2001) have shown that inhibition of MT motor proteins reduces an AMPA receptor-mediated response in hippocampal slice. This means that a labile pool of AMPA-receptors depends on MT dynamics, and MT motors determine the amplitudes of excitatory postsynaptic currents (EPSC).

     Furthermore, a walking kinesin carries with it a protein or an mRNA molecule. Since kinesin binds to a MT on a C-terminus, as it steps on it, it brings the C-terminus to the MT surface and makes it ineffective in binding for the next kinesin over a period of time that it takes the C-terminus to unbind and protrude outside. From this we can hypothesize that long stretches of C-termini in the upright position are more efficient at transporting kinesin and kinesin cargo while C-termini that lie flat may lead to a detachment of kinesin from the MT's (see Fig. 17). Thus, in considering the trafficking of many kinesins, collective conformational effects of C-termini become more and more important. While this is outside the scope of this paper, we may speculate even further and put forth a hypothesis that dendritic processing leading to axonal spikes at the axon hillock (an event not completely explained by membrane signaling) may be regulated or at least affected by the dyanamic states of the C-termini in essence leading to a complex dynamic system within a neuron

**ACKNOWLEDGMENTS**

This research was supported by grants from NSERC, MITACS-MMPD and the YeTaDel Foundation. We thank Mr. Eric Carpenter for assistance in computational work. Discussions with Dr. Dan Sackett of NIH are gratefully acknowledged.

APPENDIX A

The choice of model parameters

- The basic units in our simulation were selected as follows:
  - Mass: 1 amu = 1 Dalton = $1.66 \times 10^{-27}$ kg
  - Time: 1 ps ($1 \times 10^{-12}$ s)
  - Charge: 1 $e^-$ ($1.6 \times 10^{-19}$ C)
  - Length: 1 nm (hence, velocity is 1 nm/ps)
  - Energy: kJ/mol
  - Force: kJ/(mol · nm)
- The mass of the molecular chain of beads is estimated as 2 kD (summing up the amino acids of the C-terminus of a given human tubulin isotype), or 2000 unit masses. For simplicity we assume identical beads, i.e., for N-beads model the weight of each bead is 2kD/N. We actually take N = 11, however the first bead is totally attached to the surface.
- The net charge of the C-terminus is assumed to be –10e, again taken from the same isotype of tubulin.
- The nominal length of the C-terminus is taken to be L = 4.5nm.
- The Lennard-Jones parameters are:
  - $\varepsilon = 1$ (however recall the truncation)
  - $\sigma = 0.45$ (in nm units)
  - $R_c = 2.0^{1/6} \sigma$
- FENE parameters:
  - K = 20-30
  - $R_c = 1.5\sigma$
- Harmonic Angle force parameters:
  - $K\theta a = 40$ (the force constant)
  - $\theta_0 = 135^0$ or $3\pi/4$
- Dihedral Harmonic Angle force parameters:
  - $K\theta d = 4$ (the force constant)
  - $\varphi_0 = 0^0$ (or $\pi$)

- o    n=2 multiplication factor
- The friction coefficient used: $\Gamma=0.03$
- The noise was taken from a (3D) uniform distribution [-W..W] W=20;

Note that the friction coefficient, $\Gamma$, enters the equation of motion as part of the fluctuation-dissipation terms in the Langevin dynamics. We use the random and friction forces to represent a virtual solution without explicitly taking into account the particles in the solution. The actual parameters of the simulation are determined from the (room) temperature, the Lennard-Jones constants and the bead mass. The friction is determined as a function of the system's time unit and the time steps of the numerical integration. In our case we were interested in maintaining the higher vibrational frequencies, so the decay time was kept longer (determined from friction). The amplitude of the stochastic force then follows the Langevin constraints.

External force representation:

The external force is simulated as a traveling localized wave with a certain amplitude, as follows:
$F(z(i), t) = \text{amp} \, \text{sech}(b(z(i)-vt))^2$ where $z(i)$ is the $z$-position of the $i$th bead
Figure 8 shows the actual force on each one of the beads. The parameters of the wave were taken to be:
- $v=0.8$ [nm/ps]
- amp=6 [kJ mol$^{-1}$ nm$^{-1}$ ]
- $b=0.25$

**Figure Captions**

Fig. 1. Diagram of C-termini on the tubulin dimers of microtubules in relationship to incoming synapses. Axon terminals typically input to spines, many of which contain AMPA and NMDA glutamate receptors. Kinesin motors transport cargo along microtubules. C-termini states affect the transport of kinesin along microtubules and this affects transport of materials to active synapses.

Fig. 2. A map of the electric charge distribution on the surface of a tubulin dimer with C-termini tails present (prepared in part with MolMol (Koradi et al., 1996)). Pluses represent the areas of positive charge while minuses represent the location of negative ones. The symbols do not correspond one-to-one to elementary charge values, only to the presence of charge of a given type.

Fig. 3. A reconstruction of the surface of a tubulin dimer with two C-termini attached (based on the PDB and C-terminus sequence. The two states of the C-terminus shown are the result of MD simulation of this system.

Fig. 4. A schematic diagram of portions of two microtubules connected by a MAP2. Each portion shows the outer face of a tubulin dimer covered by two C-termini. Each C-terminus can be in one of two metastable states, perpendicular or parallel to the dimer surface.

Fig. 5. A portion of the microtubule composed of a few tubulin dimers with their C-termini used as a model for the static simulation. Also shown are a few of the positively charged regions (circles) on the surface of a tubulin dimer. The test C-terminus is slanted (bold). (Axes are scaled in [nm])

Fig. 6. Evaluation of Eq. 1 for the interaction energy between a test C-terminus and the environment vs. the azimuthal angle; each line in the figure describes an elevation cut; values are given.

Fig. 7. Energy surface of the interaction between the test C-terminus and the environment. The grayscale is in [eV]. The insert shows a zoom of a saddle point on the surface.

Fig. 8 Plot of the function $F(z(i), t) = \text{amp sech}(b(z(i)-vt))^2$ where $z(i)$ is the $z$-position of the $i$th bead.

Fig. 9. Four C-terminal conformations observed with the flexible model.

Fig. 10. The histograms shown describe the minimal energy positions of beads representing a C-terminus ranging from 3 to 5 obtained at the end of each run after relaxation. The energy minimized value can come from each of the beads at positions from 3 to 11. The first bead is permanently connected to the surface and hence has no freedom to move. The surface is taken at $z = 0.45$ nm for technical reasons.

Fig. 11. The cumulative sum of the probabilities corresponding to the same physical situation as described in Fig. 10 is just another way to present the results. It is shown that the probability of the down position which includes all cases of full or partial attachment is ~15 %.

Fig.12. A schematic model of the MAP2 interacting with positive counter-ions (circles). Also seen are two dimers at both ends of the MAP2 and a slanted C-terminus. The zoom region depicts the springs model used to describe the interactions between the MAP2 and the counter-ions.

Fig. 13. Results of the MAP2 model simulation describing the counter-ions displacement (the chain-axis denotes the discrete indices of the counter-ions). The grayscale is given in [$\text{Å}$].

Fig. 14. Time derivative of the counterions displacement evaluated using Eq. 15a. The grayscale is given in units of [$\text{Å} \cdot ps^{-1}$].

Fig. 15. Spatial derivative of the counterions displacement evaluated using Eq. 15b.

Fig 16. Evaluation of the phase velocity of the wave propagation along the MAP2 using Eq. 16

Fig. 17. C-termini that stand upright will increase kinesin processivity, whereas C-termini that lie flat will favor kinesin detachment. This could serve as a mechanism for releasing cargo near to activated synapses. The state of C-termini can be transmitted from one MT to a neighboring MT, in theory. Significant transport toward, and detachment near, active synapses could occur.

# Figures

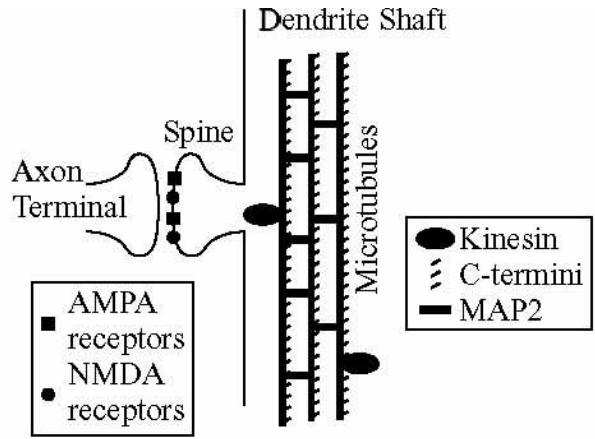

*FIGURE 1*

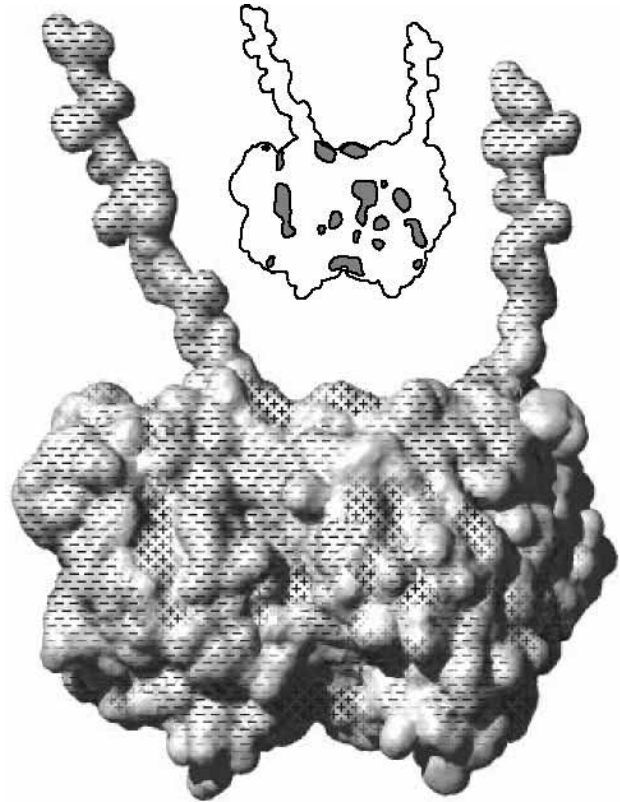

*FIGURE 2*

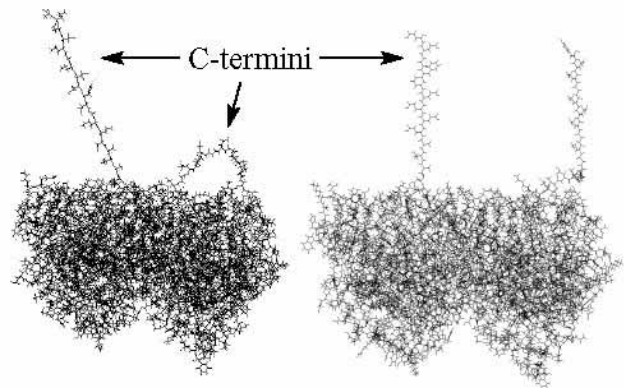

*FIGURE 3*

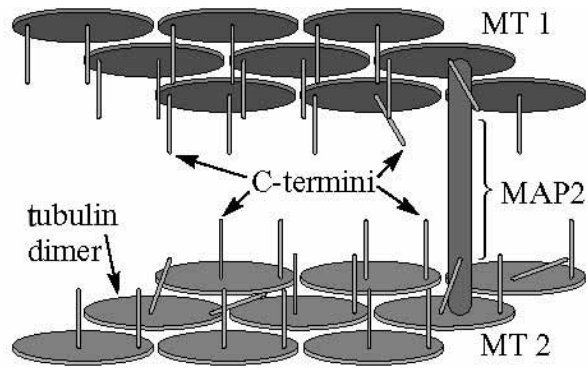

*FIGURE 4*

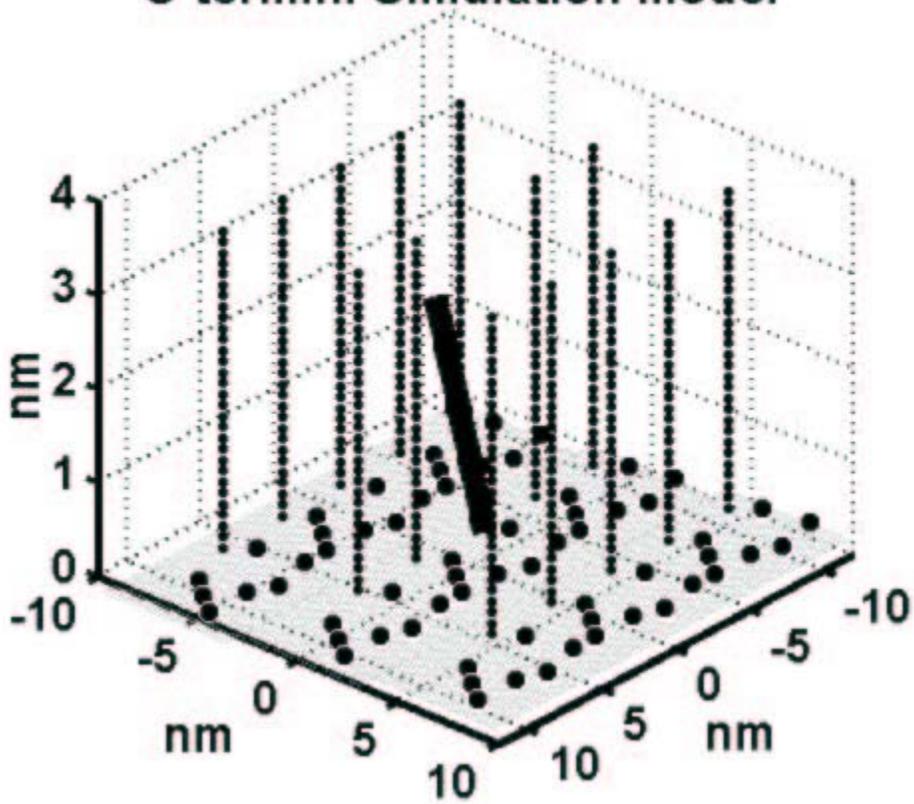

*FIGURE 5*

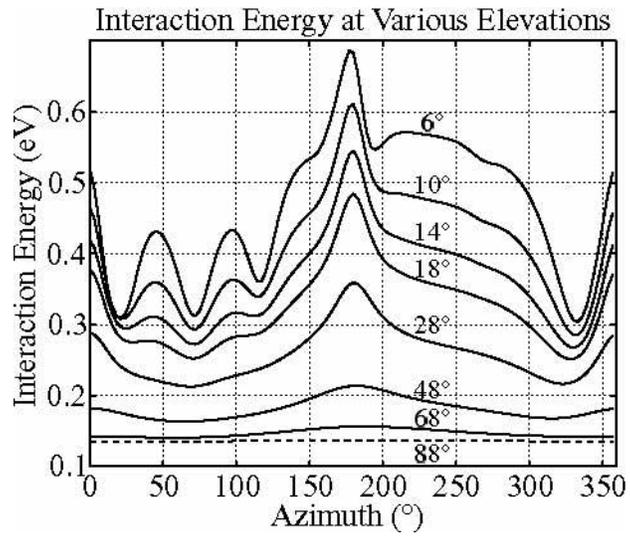

*FIGURE 6*

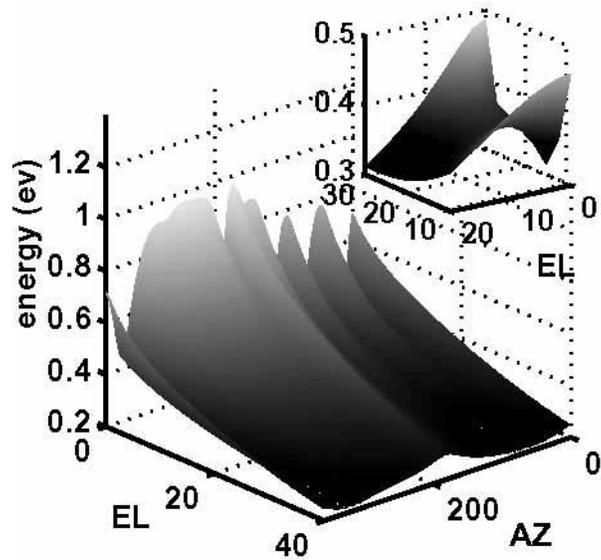
*FIGURE 7*

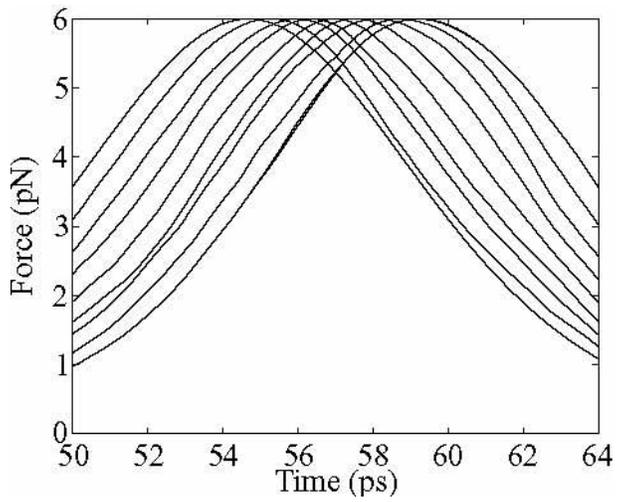
*FIGURE 8*

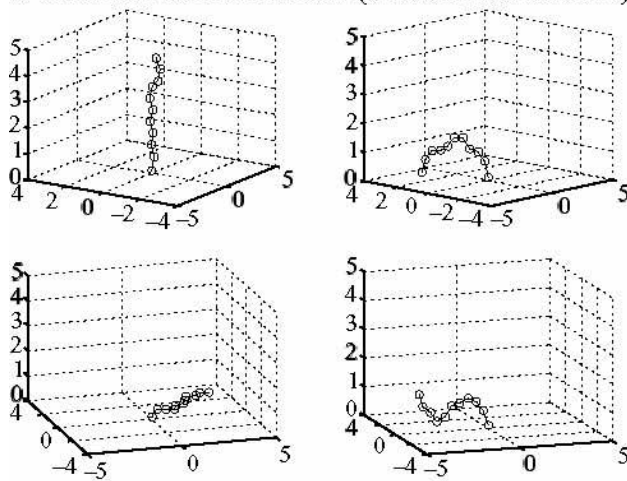

FIGURE 9

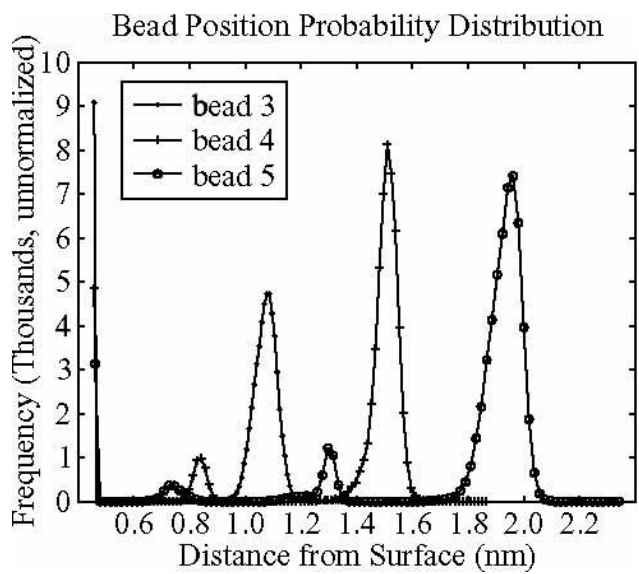

FIGURE 10

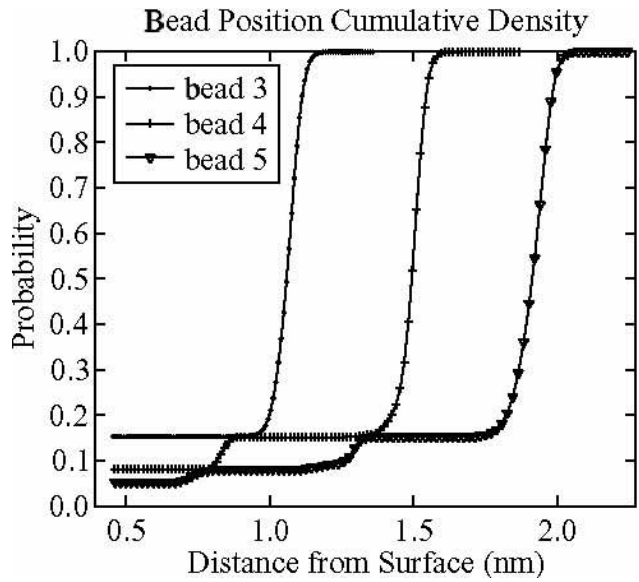

FIGURE 11

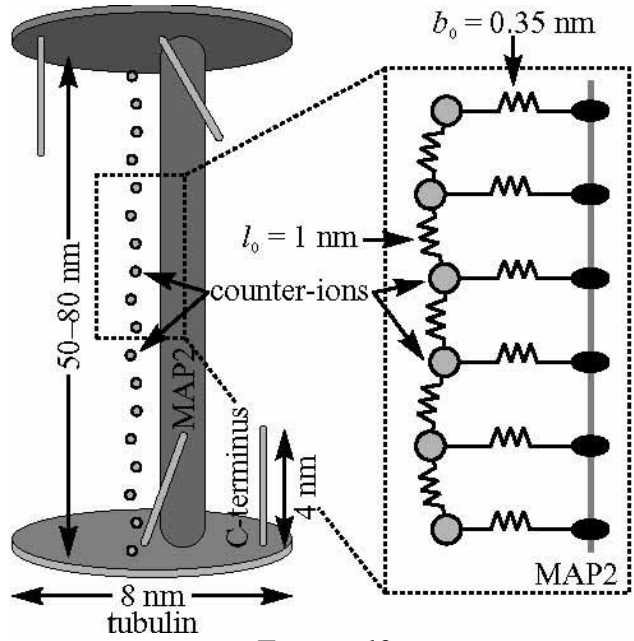

FIGURE 12

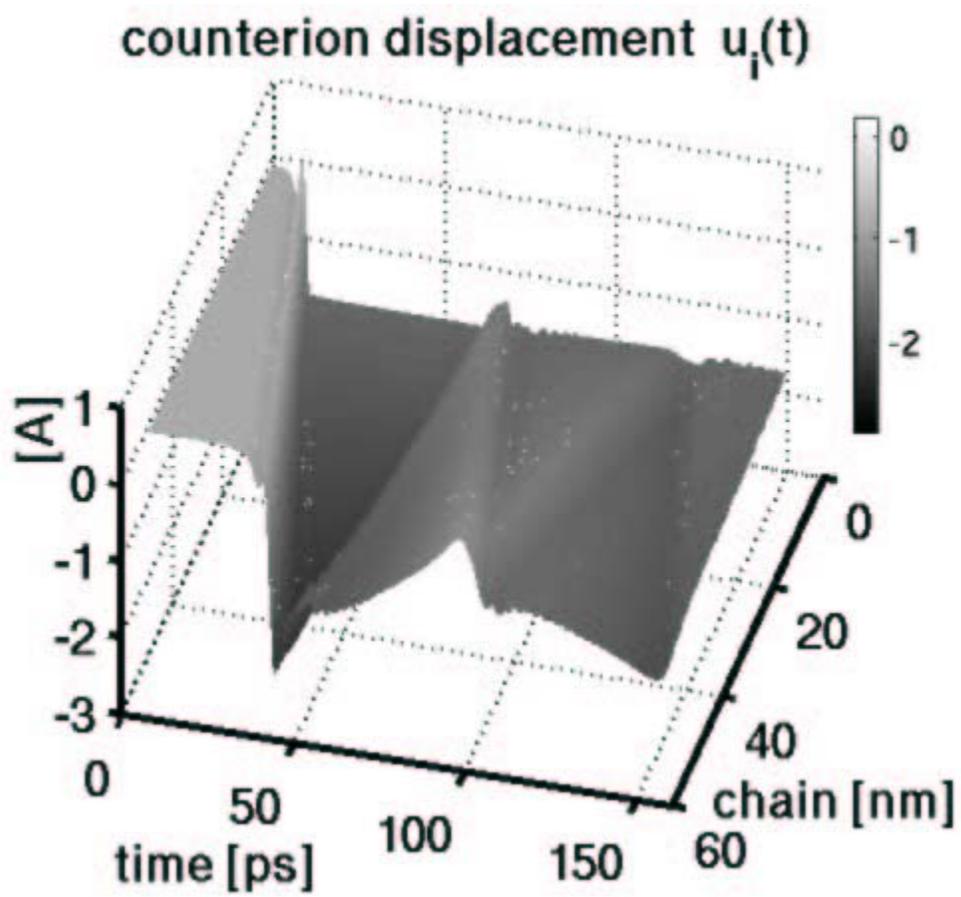

*FIGURE 13*

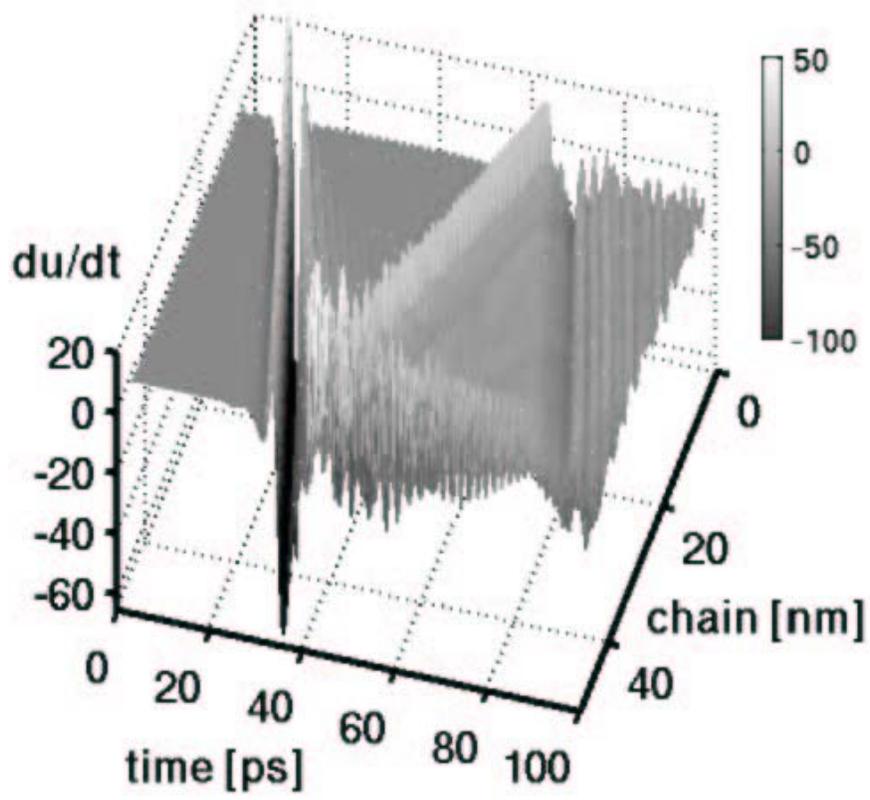

*FIGURE 14*

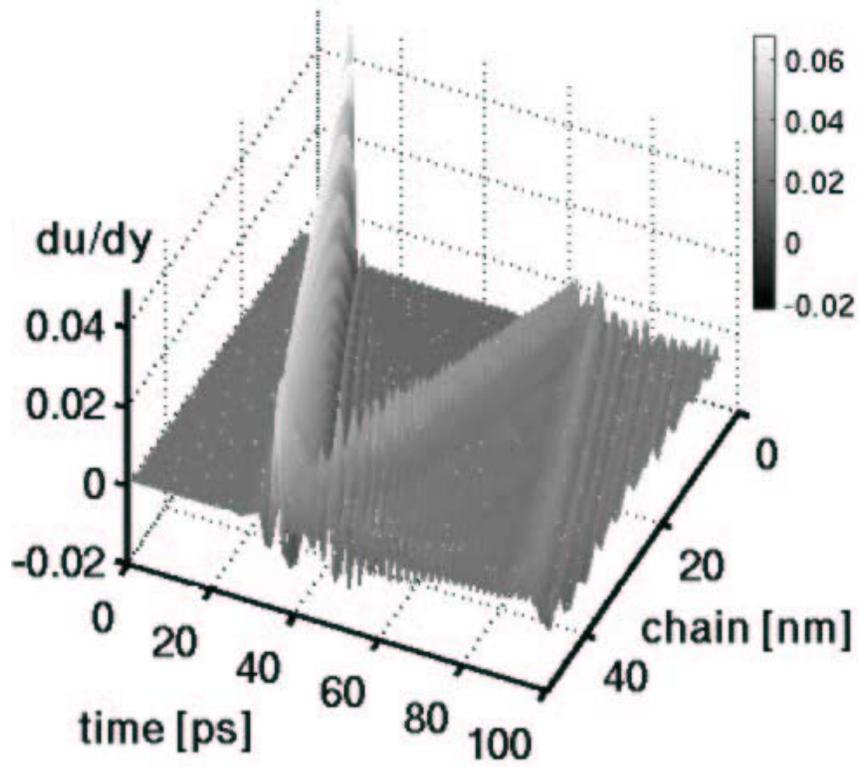

*FIGURE 15*

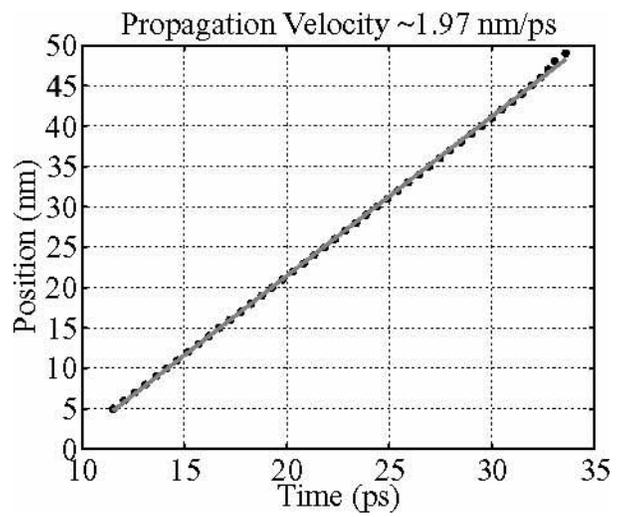
FIGURE 16

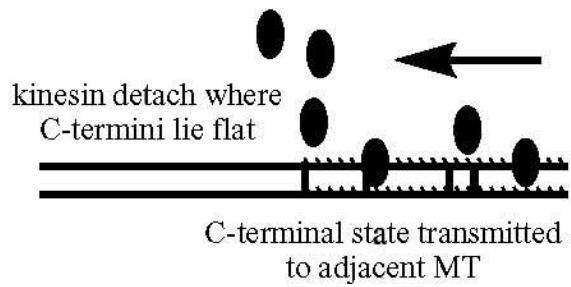
FIGURE 17